\documentclass[prl,floatfix,twocolumn,showpacs,amsmath,superscriptaddress]{revtex4-2}
\usepackage{bbm}
\usepackage{graphicx}
\usepackage{dcolumn}
\usepackage{subfigure}
\usepackage{amsmath,bm}
\usepackage{amsfonts}
\usepackage{appendix}
\usepackage{feynmf}
\usepackage{hyperref}
\usepackage{eepic}
\usepackage{lipsum}
\usepackage{enumerate}
\usepackage{amssymb}
\usepackage[english]{babel}
\usepackage{xcolor}    
 
\usepackage{multirow}

\makeatletter

\newcommand{\Rmnum}[1]{\expandafter\@slowromancap\romannumeral #1@}

\newcommand{\ua}{\uparrow} 
\newcommand{\da}{\downarrow}

\newcommand{\lra}{\leftrightarrow}
\newcommand{\bk}{\mathbf k}

\renewcommand{\vec}[1]{\mathbf{#1}}
\newcommand{\vk}{{\vec{k}}}

\begin{document}
	\title{Pair density wave facilitated by Bloch quantum geometry in nearly flat band multiorbital superconductors}
	
	\author{Weipeng Chen}
	\address{Shenzhen Institute for Quantum Science and Engineering, Southern University of Science and Technology, Shenzhen 518055, Guangdong, China}
	\address{International Quantum Academy, Shenzhen 518048, China}
	\address{Guangdong Provincial Key Laboratory of Quantum Science and Engineering, Southern University of Science and Technology, Shenzhen 518055, China}
	\author{Wen Huang}
	\email{huangw3@sustech.edu.cn}
	\address{Shenzhen Institute for Quantum Science and Engineering, Southern University of Science and Technology, Shenzhen 518055, Guangdong, China}
	\address{International Quantum Academy, Shenzhen 518048, China}
	\address{Guangdong Provincial Key Laboratory of Quantum Science and Engineering, Southern University of Science and Technology, Shenzhen 518055, China}
	
	\date{\today}
	
	\begin{abstract}
	Bloch electrons in multiorbital systems carry quantum geometric information characteristic of their wavevector-dependent interorbital mixing. The geometric nature impacts electromagnetic responses, and this effect carries over to the superconducting state, which receives a geometric contribution to the superfluid weight. In this paper, we show that this contribution could become negative under certain appropriate circumstances. This may facilitate the stabilization of Cooper pairings with real space phase modulation, i.e. the pair density wave order, as we demonstrate through two-orbital model Bogoliubov de-Gennes mean-field calculations. The quantum geometric effect therefore constitutes an intrinsic mechanism for the formation of such a novel phase of matter in the absence of external magnetic field. 

	\end{abstract}
	
	\maketitle
{\bf Introduction.--} In systems with multiple orbital degrees of freedom, electrons on the individual Bloch bands are not featureless independent particles. Instead, the motion of a Bloch electron on one band is inherently connected to that of other Bloch bands at the same wavevector, despite them being distinct energy eigenstates. More specifically, expressed in the band basis under which the kinetic part of the Hamiltonian is diagonalized, the velocity operator possesses finite off-diagonal terms, i.e. the interband velocity, 
\begin{eqnarray}\label{eq:interbandV}
	V_{\mu\vk}^{mn} =(\epsilon_{n\vk}-\epsilon_{m\vk}) \langle \psi_{m\vk}|\partial_{k_\mu} \psi_{n\vk} \rangle\,,~~~~m \neq n,
\end{eqnarray}
where $\epsilon_{m\vk}$ is the energy dispersion of the $m$-th band with eigenvector $|\psi_{m\vk} \rangle$. The object $i\langle \psi_{m\vk}|\bk\partial_{k_\mu} \psi_{n\vk} \rangle$ depicts a non-Abelian Berry connection between the Bloch states and therefore characterizes their unique quantum geometric properties. Some aspects of the geometric effect on quantum transport have been known for long. The most classic example is the quantum Hall or Chern insulator, in which the quantized Hall conductance is intimately tied to the Berry curvature of the Bloch bands~\cite{Thouless:82,Hasan:10,Qi:11}. Recently, more aspects of the geometry-induced electromagnetic responses have been discussed at length~\cite{Neupert:13,LiZ:20,Topp:21,Ahn:21}.

The geometric nature carries over to the superconducting state. Hence its footprint must also be found in the superconducting electromagnetic responses. The past several years have witnessed considerable attention on the geometry-induced finite superfluid weight in flat band superconductors~\cite{Peotta:15,Julku:16,Liang:17,Hazra:19,Verma:21,Hu:19,Julku:20,Xie:20,Torma:21,Huhtinen:22}, where conventional theory would have otherwise predicted vanishing superfluid density and hence unsustainable superconductivity. These studies are of particular relevance to the putative (near) flat band superconductivity reported in twisted bilayer graphene~\cite{Cao:18,Yankowitz:19}. Nonetheless, the geometric effect is not unique to flat band systems. Recently, the discussion has been extended to geometry-induced effects, including some peculiar optical anomalies, in a more broad spectrum of multiorbital superconductors~\cite{Chen:21,WangZQ:20,Ahn:21prb,Kitamura:22a}.  

In this paper, we demonstrate that the Bloch quantum geometry could also facilitate the formation of novel phases of matter in multiorbital superconductors. Our discussion is motivated by the observation that the geometry-related contribution to the superfluid weight is not necessarily positive definite. To be more concrete, we show that this contribution contains a part that relates to an effective interband Josephson coupling between the bands. This coupling could become negative if the superconducting order parameters on the multiple bands condense into an appropriate configuration, resulting in a suppressed phase stiffness. More strikingly, in the narrow- or flat-band limit where the pairing gap could become as large as a significant fraction of the bandwidth, the geometric contribution could increase significantly, even to the point of returning an overall negative superfluid weight. 

Since the superfluid weight characterizes the real-space superconducting phase stiffness, an intriguing question then arises: does the geometry-induced suppression of superfluid weight provide a natural mechanism for the formation of superconducting states with real-space phase mudulation, such as the Fulde-Ferrell-Larkin-Ovchinnikov (FFLO)~\cite{Fulde:64,Larkin:65} or other more general forms of pair density wave (PDW) order?  We answer this question in the affirmative, on the basis of model mean-field  Bogoliubov de-Gennes (BdG) calculations which explicitly demonstrate the preference for a PDW rather than the uniform phase. Our study therefore provides an intrinsic mechanism for PDW phases in the absence of external magnetic field.

{\bf Negative interband Josephson coupling.--} We set the stage of our discussions by first deriving the superfluid weight of a weak-coupling two-dimensional two-orbital model with uniform intraorbital spin-singlet Cooper pairing and without spin-orbit coupling. The BdG Hamiltonian matrix of the model, expressed in appropriate orbital basis, can be written as,
\begin{eqnarray} \label{eq:BdGorbital}
	\hat{H}^\text{BdG}_\vk=\begin{pmatrix}
		\hat{H}_{0\vk} & \hat{\Delta}_{\vk} \\
		\hat{\Delta}_{\vk}^\dagger & -\hat{H}^*_{0\bar{\vk}}
	\end{pmatrix} \,.
\end{eqnarray}
Here, $\bar{\vk}=-\vk$, $\hat{H}_{0\vk}$ and $\hat{\Delta}_{\vk}$ respectively describe the kinetic and pairing parts of the Hamiltonian. Following the standard linear response theory, the superfluid weight tensor can be expressed as~\cite{Liang:17}, 
\begin{eqnarray}\label{eq:Ds}
	D_{\mu \nu} &=& T_{\mu \nu} +\Pi_{\mu \nu} \nonumber \\
	&=&\frac{1}{\beta}\sum_{\omega_n,\vk}  \text{Tr} \{ \hat{m}^{-1}_{\mu \nu} \hat{\mathcal{G}}_\vk(i\omega_n)+\hat{\mathcal{G}}_\vk(i\omega_n) \hat{V}_{\mu\vk} \hat{\mathcal{G}}_\vk(i\omega_n)\hat{V}_{\nu\vk}\} \,, \nonumber \\
	&&
\end{eqnarray}
where $\mu,\nu=x,y$, $\omega_n=(2n+1)\pi/\beta$ denotes the fermionic Matsubara frequency, and $\hat{\mathcal{G}}_{\vk}(i\omega_n)= (i\omega_n - \hat{H}^\text{BdG}_\vk)^{-1}$ is the Gor'kov Green's function. The velocity operator $\hat{V}_{\mu\vk}$ takes the following form, 
\begin{eqnarray}\label{eq:Vorb}
	\hat{V}_{\mu\vk} = \tau_z \partial_{k_\mu}\hat{H}^\text{BdG}_\vk \big|_{\Delta \rightarrow 0} \,,
\end{eqnarray}
where $\tau_z$ is the third component of the Pauli matrix operating in the particle-hole space of the BdG Hamiltonian (\ref{eq:BdGorbital}). 

The second term in (\ref{eq:Ds}) represents the paramagnetic contribution, which vanishes in single-orbital models at zero temperature. The first term, $T_{\mu\nu}$, is the diamagnetic contribution in which the inverse mass tensor is given by,
\begin{eqnarray}\label{eq:invM}
	\hat{m}^{-1}_{\mu\nu}=\frac{\partial^2\hat{H}^\text{BdG}_\vk}{\partial {k_\mu} \partial {k_\nu}} \Big|_{\Delta \rightarrow 0} \,.
\end{eqnarray}
Utilizing (\ref{eq:invM}) and the periodic boundary condition in the Brillouin zone of a lattice model, the diamagnetic term can be written in a different form,
\begin{eqnarray}\label{eq:T}
	T_{\mu\nu} &=&-\frac{1}{\beta}\sum_{\omega_n,\vk}  \text{Tr} \left[\hat{\mathcal{G}}_\vk \partial_{k_{\mu}}\hat{H}^\text{BdG}_\vk \hat{\mathcal{G}}_\vk \left(\partial_{k_{\nu}}\hat{H}^\text{BdG}_\vk\big|_{\Delta \rightarrow 0}\right)  \right] \nonumber \\
	&\approx&-\frac{1}{\beta}\sum_{\omega_n,\vk}  \text{Tr} \left[\hat{\mathcal{G}}_\vk \tau_z \hat{V}_{\mu\vk} \hat{\mathcal{G}}_\vk \tau_z \hat{V}_{\nu\vk}  \right]   \,.
\end{eqnarray}
In the second line of the above equation, we have taken the approximation, $\partial_{k_{\mu}}\hat{H}^\text{BdG}_\vk \approx \partial_{k_{\mu}}\hat{H}^\text{BdG}_\vk\big|_{\Delta \rightarrow 0}  = \tau_z \hat{V}_{\mu\vk}$, which is valid in the weak-coupling limit as terms with $\partial_{k_\mu}\hat{\Delta}_{\vk}$ are expected to be much smaller.

\begin{figure}
	\centering
	\includegraphics[scale=0.32]{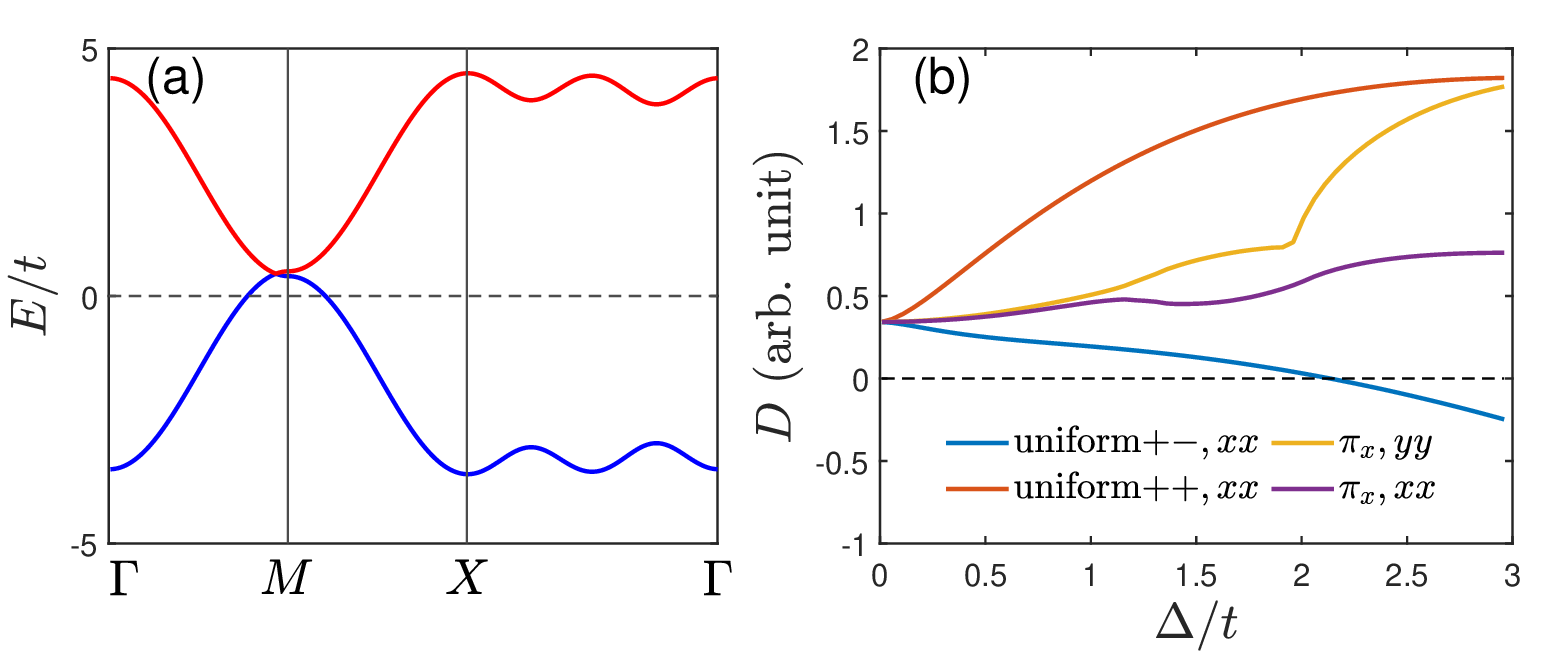}
	\caption{(color online) (a) The band structure of the $s$-$d_{xy}$ lattice model used in our calculation. (b) The superfuid weight $D_{xx/yy}$ as a function of the intraorbital pairing potential $\Delta$. The blue (red) curve denotes superfluid weight of the uniform phase with order parameter configuration $\hat{\Delta}_{+-}$ ($\hat{\Delta}_{++}$). The purple and yellow curves respectively display the $xx$ and $yy$ components of the superfluid weight in the $\pi$-PDW order associated with $\hat{\Delta}_{+-}$. The tight-binding parameters are set to $(t_s, t_d, \mu_s,\mu_d, t')=(1, -1, -0.5, -0.4, 1) t$.
	}
	\label{fig01}
\end{figure}

To facilitate our subsequent derivations, we turn to the band basis representation, using a unitary transformation that diagonalizes the kinetic part of (\ref{eq:BdGorbital}), i.e. $\hat{U}^{-1}_\vk \hat{H}_\vk^\text{BdG}\hat{U}_\vk = \tilde{H}^\text{BdG}_\vk$. With a further assumption that the weak Cooper pairing takes place only between electrons on the same band, i.e. intraband pairing, the BdG Hamiltonian in the basis $(c_{1\vk\ua},c_{2\vk\ua},c^\dagger_{1\bar{\vk}\da},c^\dagger_{2\bar{\vk}\da})^T$ reads,
\begin{eqnarray} \label{eq:Hbdg}
	\tilde{H}^\text{BdG}_\vk=\begin{pmatrix}
		\epsilon_{1\vk} && \Delta_{1\vk} &\\
		&\epsilon_{2\vk}&& \Delta_{2\vk}  \\
		\Delta^*_{1\vk} &&  -\epsilon_{1\vk}& \\
		&\Delta^*_{2\vk}& &-\epsilon_{2\vk}
	\end{pmatrix}\,.
\end{eqnarray}
Here, $\epsilon_{m\vk}$ and $\Delta_{m\vk}$ give, respectively, the normal state dispersion and the intraband pairing function of band-$m$, where $m=1,2$ label the band indices. For future reference, we write down the resultant Bogoliubov dispersion $E_{m\vk}= \sqrt{\epsilon^2_{m\vk}+|\Delta_{m\vk}|^2}$ with corresponding eigenstate $|m\vk\rangle=(u_{m\vk},v_{m\vk})^T=(\Delta_{m\vk},E_{m\vk}-\epsilon_{m\vk})^T/N_{m\vk}$ and its particle-hole symmetric state $|\bar{m}\vk\rangle=(-v^\ast_{m\vk},u^\ast_{m\vk})^T$, where $N_{m\vk}$ is a normalization factor. The velocity operator in the same basis can be obtained following the same unitary transformation,
\begin{eqnarray}\label{eq:Vband}
	\tilde{V}_{\mu\vk} =\begin{pmatrix}
		V^{11}_{\mu\vk} & V^{12}_{\mu\vk} &  &  \\
		V^{21}_{\mu\vk} & V^{22}_{\mu\vk} &  &  \\
		& & -(V^{11}_{\mu\bar{\vk}})^\ast & -(V^{12}_{\mu\bar{\vk}})^\ast \\
		&  & -(V^{21}_{\mu\bar{\vk}})^\ast & -(V^{22}_{\mu\bar{\vk}})^\ast  
	\end{pmatrix},
\end{eqnarray} 
where the diagonal elements are simply the band group velocities $V^{mm}_{\mu\vk}=-V^{mm}_{\mu\bar{\vk}}=-(V^{mm}_{\mu\bar{\vk}})^\ast=\partial _{k_\mu}\epsilon_{m\vk}$, while the off-diagonal elements, i.e. the interband velocities, as given in (\ref{eq:interbandV})~\cite{Liang:17,Chen:21}, satisfy $V^{12}_{\mu\vk}=(V^{21}_{\mu\vk})^\ast$. Finally, the superfluid weight (\ref{eq:Ds}), having the diamagnetic term substituted by (\ref{eq:T}), is obtained by replacing the velocity operator with (\ref{eq:Vband}) and the Green's function with $\tilde{\mathcal{G}}_{\vk}(i\omega_n) = (i\omega_n-\tilde{H}^\text{BdG}_\vk)^{-1} = \sum_{m} \left( \frac{|m\vk \rangle\langle m\vk| }{i\omega_n - E_{m\vk}} + \frac{|\bar{m}\vk \rangle\langle \bar{m}\vk| }{i\omega_n + E_{m\vk}} \right)$. The procedure is equivalent to inserting identity operators $\hat{I}= \hat{U}^{-1}_\vk \hat{U}_\vk$ into (\ref{eq:Ds}) and then dropping the interband pairings. Taking the $xx$-component as an explicit example, the zero-temperature superfluid weight reads as follows,
\begin{eqnarray}
	D_{xx} &=& 4\sum_m \frac{\big|V_{x\vk}^{mm} u_{m\vk}v_{m\vk}\big|^2}{E_{m\vk}} \nonumber \\
		&&-16\text{Re}\left[\frac{V^{12}_{x\vk}(V^{21}_{x\bar{\bk}})^\ast u^*_{1\vk}v_{1\vk}u_{2\vk}v^*_{2\vk}}{E_{1\vk}+E_{2\vk}} \right].
	\label{eq:rho}
\end{eqnarray}
As is shown in more detail in the Supplementary, there are multiple terms from the diamagnetic and paramagnetic responses that exactly cancel each other. Hence the two terms on the rhs of this equation are not in one to one correspondence with the two respective responses. Nonetheless, the first term, generated by the virtual intraband transitions $| \bar{m}\vk\rangle \lra |m\vk\rangle$, originates solely from the diamagnetic contribution. In the weak-coupling and isolated-band limit, it returns the conventional expression for the zero-temperature superfluid weight. On the other hand, the second term receives equal contributions from the diamagnetic and paramagnetic responses. Absent in traditional theories, this term is purely of quantum geometric origin. In particular, it is associated exclusively with virtual interband transitions such as $|\bar{2}\vk\rangle \lra |1\vk\rangle$, thereby acquiring an explicit dependence on the interband velocity. 

Note that, in this approximation the geometric term is finite only when both bands develop Cooper pairing. In fact, it can be cast in a more suggestive form $V^{12}_{x\vk}(V^{21}_{x\bar{\bk}})^\ast  \langle c^\dagger_{1\vk\ua}c^\dagger_{1\bar{\vk}\da} \rangle \langle c_{2\bar{\vk}\da}c_{2\vk\ua} \rangle$, where $\langle \cdots \rangle $ denotes the expectation value of the ground state. It is thus indicative of an effective interband Josephson coupling, where the interband velocity serves to `tunnel' Cooper pairs from one band to another. Written more explicitly, this term reads, 
\begin{eqnarray}
	-16\text{Re}\left[V^{12}_{x\vk}(V^{21}_{x\bar{\bk}})^\ast \frac{\Delta^*_{1\vk}\Delta_{2\vk}(E_{1\vk}-\epsilon_{1\vk})(E_{2\vk}-\epsilon_{2\vk})}{N^2_{1\vk}N^2_{2\vk}(E_{1\vk}+E_{2\vk})}\right].
	\label{eq:Geometric}
\end{eqnarray}
An interesting observation is that, unlike the first term in (\ref{eq:Ds}) which is positive definite, this term could become negative in certain scenarios. In a later illustrative lattice model, $V^{12}_{\mu\vk}$ is odd in $\vk$, and it could be real under a proper gauge choice. Taking a simple example where $\Delta_{m\vk} \equiv \Delta_m$, the geometric correction (\ref{eq:Geometric}) is thus negative if $\text{sign}[\Delta_1] = -\text{sign}[\Delta_2]$. In this case, the total superfluid weight is reduced from the approximation based on conventional isolated-band considerations. Conversely, the superfluid weight is enhanced if the two gaps are of the same sign. Note that the sign of the geometric term must be analyzed on a case by case basis, because $V^{12}_{x\vk}$ could be even in $\vk$ and $\Delta_{m\vk}$ may acquire wavevector dependence in different models. 

The nontrivial geometric contribution to the superfluid weight has been discussed in quite a number of previous studies~\cite{Peotta:15,Julku:16,Liang:17,Hazra:19,Verma:21,Hu:19,Julku:20,Xie:20,Torma:21,Huhtinen:22}. In general, the correction increases with increasing pairing strength~\cite{Julku:16}. However, in weak-coupling limit the coherence factor $|u_{m\vk}v_{m\vk}|$ is peaked around the Fermi wavevectors of band-$m$ and vanishes elsewhere. Hence the term (\ref{eq:Geometric}) is negligible, except in rare scenarios where the two Fermi surfaces overlap. A bold conjecture is that a strong geometric contribution could be obtained in a model with large $\Delta/E_F$, which would see a significantly broadened distribution of $|u_{m\vk}v_{m\vk}|$. However, in this case the approximation made in (\ref{eq:T}) is no longer valid, and interband pairing may also come into play. Below, we turn to an orbital-basis analysis and extend to the regime where $\Delta$ is comparable to the Fermi energy or the bandwidth. We shall still employ the BdG formalism and the mean-field linear-response theory for a first-stage study in search for some preliminary qualitative understanding.
	
{\bf Strong geometry-induced suppression of $D$.}-- The possibility of negative superfluid weight has in fact been noted in a previous study on twisted bilayer graphene~\cite{Xie:20}, although the implication for PDW order was not mentioned there. Here, we explicitly demonstrate negative superfluid weight in a simpler and more mundane two-orbital model consisting of an $s$- and a $d_{xy}$-orbital on a square lattice, although the conclusion applies to more general models. Assuming only onsite intraorbital spin-singlet pairing on the two orbitals, the BdG Hamiltonian matrix can be written in the basis $(c_{s\vk\ua},c_{d\vk\ua},c^\dagger_{s\bar{\vk}\da},c^\dagger_{d\bar{\vk}\da})^T$ as, 
\begin{eqnarray}		
	\hat{H}^\text{BdG}_\vk =\begin{pmatrix}
		\xi_{s\vk} & \lambda_\vk & \Delta_s  & 0 \\
		\lambda^*_\vk & \xi_{d\vk} & 0  & \Delta_d \\
		\Delta_s^\ast & 0 & -\xi_{s\bar{\vk}} & -\lambda^\ast_{\bar{\vk}} \\
		0 & \Delta_d^\ast & -\lambda_{\bar{\vk}} & -\xi_{d\bar{\vk}} 
	\end{pmatrix}\,,
\end{eqnarray}
where $\xi_{a\vk}=-2t_a(\cos k_x +\cos k_y)-\mu_a$ $(a=s,d)$ represent the intraorbital dispersion and $\lambda_\vk=4t^\prime\sin k_x \sin k_y$ denotes the interorbital mixing. In practice, we keep a balance between $t^\prime$ and $t_s-t_d$, $\mu_s-\mu_d$, so as to ensure a sizable interband velocity. At this stage, we assume $|\Delta_s| = |\Delta_d| = \Delta$ which, when transformed into the band basis, happens to determine the intraband pairing order parameter configuration on the two bands (see the Supplementary). Specifically, the $++$ configuration with $\hat{\Delta}_{++}=(\Delta_s, \Delta_d) = (\Delta, \Delta)$ leads to $\Delta_{1\vk} = \Delta_{2\vk}$, while the $+-$ configuration with $\hat{\Delta}_{+-} = (\Delta,-\Delta)$ gives rise to $\Delta_{1\vk} = -\Delta_{2\vk}$. Hence, according to (\ref{eq:Geometric}), these two distinct uniform phases likely carry opposite geometric corrections to the superfluid weight. 

The superfluid weight obtained from (\ref{eq:Ds}) for a representative given band structure is shown in Fig.~\ref{fig01}. Consistent with (\ref{eq:Geometric}), a dichotomy between the $\hat{\Delta}_{++}$ and $\hat{\Delta}_{+-}$ models is indeed observed, with the latter exhibiting a monotonically decreasing $D_{xx}$ as a function of the gap amplitude. The most striking scenario with negative $D$ in the $+-$ model is achieved when $\Delta$ exceeds a large model-dependent value. Since the conventional contribution is positive definite, the negative superfluid weight must be attributed to the quantum geometric effects alone. 

\begin{figure}
	\centering
	\includegraphics[scale=0.5]{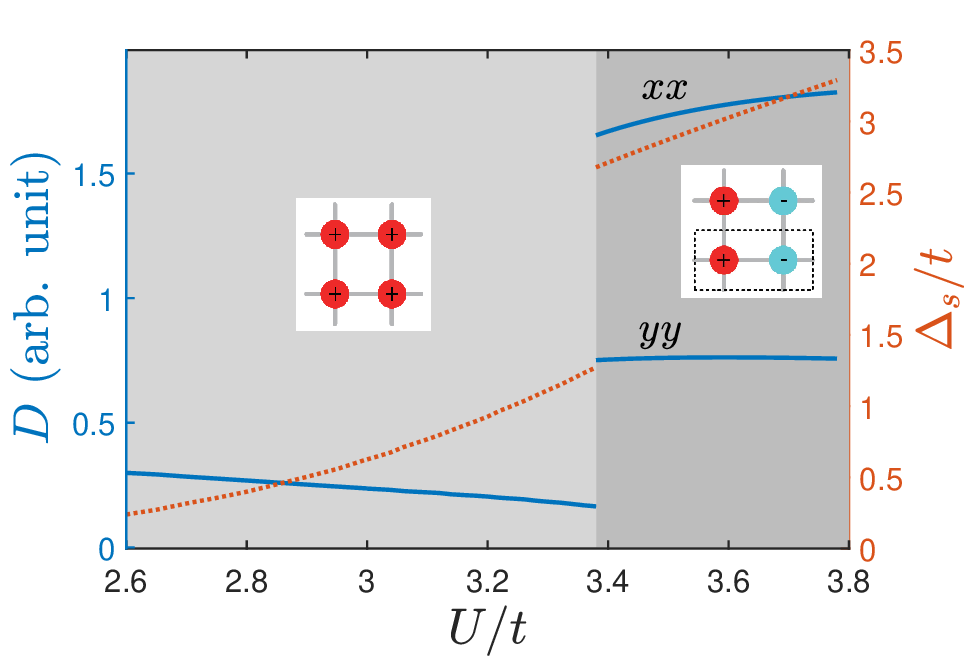}
	\vspace{-0.cm}
	\caption{(color online) The superconducting phase diagram as a function of the pairing interaction. A transition between uniform and $\pi$-PDW phases occurs at about $U/t=3.38$. The blue solid curves represent the superfluid weight $D_{xx/yy}$, while the orange dotted lines indicate the amplitude of $\Delta_s$, i.e. the intraorbital pairing on the $s$-orbital. The insets sketch the uniform and $\pi$-PDW orders on the square lattice. The ``$+/-$'' signs on the sites indicate the overall phase modulation of the pairing potential $\hat{\Delta}_{+-}$, while the black dashed rectangle depicts the unitcell of $\pi$-PDW.
	}
	\label{fig03}
\end{figure}

Showing a negative $D$ for the uniform phase is insufficient to judge the plausibility of any PDW order. To this end, we introduce a new superfluid weight defined on the basis of the corresponding enlarged unitcell of the PDW order. Similar to its original definition, such a quantity measures the phase stiffness among the enlarged unitcells. We take the $\hat{\Delta}_{+-}$ configuration and, for illustrative purpose, consider what we call a $\pi$-PDW --- one that has modulation wavevector $(\pi,0)$ and sees the pairing potential change sign every other site in the $x$-direction (illustrated in the inset of Fig.~\ref{fig03}). The calculation will adopt a new velocity operator expressed in the enlarged basis with two lattice sites per unitcell. Details are provided in the Supplementary. 

The superfluid weight $D_{xx}$ and $D_{yy}$ defined for $\pi$-PDW as a function of $\Delta$ are also plotted in Fig.~\ref{fig01} (b) for comparison. The two components differ because the $\pi$-PDW breaks the four-fold rotational symmetry. Note that the $xx$-component exhibits an apparent cusp at around $\Delta = 2t$. In close inspection, this coincides with a qualitative change in the quasiparticle spectrum, i.e. going from having a Bogoliubov Fermi surface~\cite{Agterberg:2017} to being fully-gapped. Intriguingly enough, both of the two components remain positive even when the uniform superfluid weight turns negative. This suggests, at the very least, that a $\pi$-PDW phase has the potential to materialize as real physical order as the uniform phase becomes unstable towards phase decoherence. 

{\bf Stabilization of PDW phase.}-- The above analyses is based on the assumption that the PDW pairing potential develops the same magnitude as the uniform phase. It is so far unclear whether the PDW could become more favorable under a given set of pairing interactions. In the following, we self-consistently determine the relative stability between the uniform and $\pi$-PDW phases. Note that the actual ground state could well be a more complex form of PDW with a different modulation wavevector. However, finding the ground state is not our primary objective here. It suffices to demonstrate the preference of at least one certain PDW order over the uniform phase. 

We consider onsite pairing interactions $H_\text{int} = \sum_i H_{\text{int},i}$, where on each site $i$, 
\begin{eqnarray}
	H_{\text{int},i} &=& -U_{ss} c^\dagger_{i,s\ua}c^\dagger_{i,s\da} c_{i,s\da}c_{i,s\ua} -U_{dd} c^\dagger_{i,d\ua}c^\dagger_{i,d\da} c_{i,d\da}c_{i,d\ua} \nonumber \\
	&& + U_{sd} \left[ c^\dagger_{i,s\ua}c^\dagger_{i,s\da} c_{i,d\da}c_{i,d\ua}  + (s \leftrightarrow d)\right].
	\label{eq:interactions}
\end{eqnarray}
Here, $U_{ss}, U_{dd}, U_{sd}>0$ designate the strength of the intraorbital ($ss$ and $dd$) interactions and the interorbital ($sd$) pair hopping. Among them, $U_{ss}$ and $U_{dd}$ promote onsite intraorbital spin-singlet pairings $\Delta_s$ and $\Delta_d$ within the two respective orbital manifolds. A sizable interorbital repulsion $U_{sd}$ is chosen to ensure that $\Delta_s$ and $\Delta_d$ condense into the $\hat{\Delta}_{+-}$ configuration. The order parameters $\Delta_s$ and $\Delta_d$ are obtained by self-consistently solving the gap equations at zero-temperature. The calculations are performed using the same tight-binding model as in Fig.~\ref{fig01}, but with a unitcell stretching two neighboring lattice sites in the $x$-direction to accommodate the $\pi$-PDW (see Fig.~\ref{fig03}). Throughout the calculation, the electron filling is kept roughly constant by properly adjusting the chemical potential. Owing to the large interorbital mixing $t^\prime$, the pairings of the two orbitals are strongly coupled, which results in comparable $|\Delta_s|$ and $|\Delta_d|$ (not shown in Fig.~\ref{fig03}). Thus the superfluid weight of a uniform or PDW phase with a self-consistently obtained pairing amplitude could also be roughly estimated by referring back to Fig.~\ref{fig01} (b). 

Keeping the relative strength of the interactions at $(U_{ss},U_{dd},U_{sd})=(1,0.95,1.2)U$, Fig.~\ref{fig03} plots the phase diagram of the model as a function of $U$. Generally speaking, the pairing becomes progressively stronger as the interaction increases. In the mean time, the uniform phase features increasingly reduced superfluid weight, in agreement with the preceding analyses. The $\pi$-PDW phase sets in beyond a critical interaction strength. Notably, the transition occurs before the uniform phase reaches negative superfluid stiffness. This can be attributed to the much stronger pairing developed in the PDW than in the uniform phase, as is shown in Fig.~\ref{fig03}. We note that the PDW phase features positive superfluid weight throughout the concerned range of interaction strength. On the other hand, no transition to the PDW phase has been found in the $\hat{\Delta}_{++}$ configuration (favored for small or negative $U_{sd}$). This further testifies how the negative geometric superfluid weight facilitates the formation of PDW order.


{\bf Conclusions.}-- PDW states have received much interest in connection to the pesudogap phase in cuprate superconductors~\cite{Fradkin:15,Agterberg:20}. There, strong Coulomb correlations are widely thought to play a pivotal role, although the exact mechanism by which they drive PDW remains an open question. PDW order has also been proposed in other strongly correlated systems, such as doped $U(1)$ quantum spin liquid~\cite{Yang:21}. The FFLO states proposed even earlier~\cite{Fulde:64,Larkin:65} do not rely on strong correlations, but require a finite Zeeman field such that the weak Cooper pairing develops a center of mass momentum. In this study, we have provided a new intrinsic mechanism for the stabilization of PDW phases in the absence of external magnetic field, in nearly flat band multiorbital superconductors. We argued that, under appropriate circumstance, the superfluid weight may receive a large negative quantum geometric contribution, which promotes the formation of Cooper pairing with real space phase modulation, i.e.~the PDW order. Our study opens a new route in the search for PDW phases in superconductors, and may be particularly relevant to the superconductivity in twisted bilayer graphene and the like. Yet to be explored, the Bloch quantum geometry may have the potential to empower other exotic phases of matter. However, caution is needed as our study has relied on the mean-field BCS approach to what is essentially a strong coupling superconductor. It is unclear what a more serious theoretical technique more suitable for strong coupling superconductors could reveal to us. This will be an interesting direction to further investigate.

{\bf Note added.} In the process of completing this manuscript, we became aware of a study which discusses the impact of quantum geometry on the FFLO states in the presence of external magnetic field~\cite{Kitamura:22}. However, the authors did not discuss the intrinsic formation of PDW order which occurs without the influence of external field. After our work was initially posted as arXiv: 2208.02285, another work with
similar idea appeared~\cite{Jiang:23}.

{\bf Acknowledgements} We acknowledge helpful discussions with Shuai Chen, Wen Sun and Zhiqiang Wang. This work is supported by NSFC under grant No.~11904155, the Guangdong Provincial Key Laboratory under Grant No.~2019B121203002, the Guangdong Science and Technology Department under Grant 2022A1515011948, and a Shenzhen Science and Technology Program (Grant No. KQTD20200820113010023). Computing resources are provided by the Center for Computational Science and Engineering at Southern University of Science and Technology.

\onecolumngrid
\vspace{\columnsep}
\section*{supplemental materials for `` Pair density wave facilitated by Bloch quantum geometry in nearly flat band multiorbital superconductors "}
\renewcommand{\theequation}{S\arabic{equation}}
\setcounter{equation}{0}
\subsection*{\label{sec:level1} Diamagnetic and paramagnetic contributions}
Following the derivation in the maintext, the diamagnetic and paramagnetic contributions to the superfluid weight can be approximated by replacing the velocity operator and the Green's function in Eq.~(\ref{eq:Ds}) in the maintext by their band basis counterparts. It is straightforward to obtain the following,
\begin{eqnarray}\label{eqn14}	
	T_{\mu \nu}&=&4\sum_m \frac{V_{\mu\vk}^{mm}V_{\nu\vk}^{mm}|u_{m\vk}v_{m\vk}|^2}{E_{m\vk}}\nonumber \\
	&&+\frac{2(V^{21}_{\mu\vk}V^{12}_{\nu\vk}|u_{1\vk}v_{2\vk}|^2+(V^{21}_{\mu\bar{\vk}})^*(V^{12}_{\nu\bar{\vk}})^*|u_{2\vk}v_{1\vk}|^2-(V^{21}_{\mu\bar{\vk}})^*V^{12}_{\nu\vk}u^*_{1\vk}v_{1\vk}u_{2\vk}v^*_{2\vk}-V^{21}_{\mu\vk}(V^{12}_{\nu\bar{\vk}})^*u_{1\vk}v^*_{1\vk}u^*_{2\vk}v_{2\vk})+(\mu \leftrightarrow\nu)}{E_{1\vk}+E_{2\vk}} \nonumber \\
	&=&4\sum_m \frac{V_{\mu\vk}^{mm}V_{\nu\vk}^{mm}|u_{m\vk}v_{m\vk}|^2}{E_{m\vk}}\nonumber \\ &&+\frac{2(V^{21}_{\mu\vk}V^{12}_{\nu\vk}|u_{1\vk}v_{2\vk}|^2+V^{12}_{\mu\bar{\vk}}V^{21}_{\nu\bar{\vk}}|u_{2\vk}v_{1\vk}|^2-V^{12}_{\mu\bar{\vk}}V^{12}_{\nu\vk}u^*_{1\vk}v_{1\vk}u_{2\vk}v^*_{2\vk}-V^{21}_{\mu\vk}V^{21}_{\nu\bar{\vk}}u_{1\vk}v^*_{1\vk}u^*_{2\vk}v_{2\vk})+(\mu \leftrightarrow\nu)}{E_{1\vk}+E_{2\vk}},
	\\
	\Pi_{\mu \nu}&=&-\frac{2(V^{21}_{\mu\vk}V^{12}_{\nu\vk}|u_{1\vk}v_{2\vk}|^2+(V^{21}_{\mu\bar{\vk}})^*(V^{12}_{\nu\bar{\vk}})^*|u_{2\vk}v_{1\vk}|^2+(V^{21}_{\mu\bar{\vk}})^*V^{12}_{\nu\vk}u^*_{1\vk}v_{1\vk}u_{2\vk}v^*_{2\vk}+V^{21}_{\mu\vk}(V^{12}_{\nu\bar{\vk}})^*u_{1\vk}v^*_{1\vk}u^*_{2\vk}v_{2\vk})+(\mu \leftrightarrow\nu)}{E_{1\vk}+E_{2\vk}} \nonumber \\
	&=&-\frac{2(V^{21}_{\mu\vk}V^{12}_{\nu\vk}|u_{1\vk}v_{2\vk}|^2+V^{12}_{\mu\bar{\vk}}V^{21}_{\nu\bar{\vk}}|u_{2\vk}v_{1\vk}|^2+V^{12}_{\mu\bar{\vk}}V^{12}_{\nu\vk}u^*_{1\vk}v_{1\vk}u_{2\vk}v^*_{2\vk}+V^{21}_{\mu\vk}V^{21}_{\nu\bar{\vk}}u_{1\vk}v^*_{1\vk}u^*_{2\vk}v_{2\vk})+(\mu \leftrightarrow \nu)}{E_{1\vk}+E_{2\vk}} \,
	.\end{eqnarray}
Here we make use of the relations $V^{mm}_{\mu\bar{\vk}}=-V^{mm}_{\mu\vk}$ and $(V^{mn}_{\mu\vk})^*=V^{nm}_{\mu\vk}$. Note that there are terms from the two contributions that are opposite to each other. Taken together, the total superfluid weight follows as,
\begin{eqnarray}
	D_{\mu\nu}=T_{\mu \nu} +\Pi_{\mu \nu}=4\sum_m \frac{V_{\mu\vk}^{mm}V_{\nu\vk}^{mm}|u_{m\vk}v_{m\vk}|^2}{E_{m\vk}}+8\text{Re}\left[\frac{V^{12}_{\mu\vk}V^{12}_{\nu\vk}u^*_{1\vk}v_{1\vk}u_{2\vk}v^*_{2\vk}+(\mu \leftrightarrow \nu)}{E_{1\vk}+E_{2\vk}}\right]\,.
\end{eqnarray}
Here we use the fact in our $s$-$d_{xy}$ model that $V^{mn}_{\mu\bar{\vk}}=-V^{mn}_{\mu\vk}$ for $m\neq n$.	 

\subsection*{\label{sec:level1} Orbital- to band-basis transformation}
The kinetic part of the Hamiltonian of our two-orbital model in the basis $(c_{s\vk},c_{d\vk})^T$ is given by
\begin{eqnarray}
	\hat{H}_{0,\vk}=\begin{pmatrix}
		\xi_{s\vk} & \lambda_\vk \\
		\lambda^*_\vk &\xi_{d\vk}
	\end{pmatrix}
	.\end{eqnarray}
This Hamiltonian can be diagonalized by the following unitary matrix,
\begin{eqnarray}\centering
	\hat{U}_\vk= \begin{pmatrix}
		\lambda_{\vk} & \xi_{\vk}-\epsilon_{\vk}  \\
		\epsilon_{\vk}-\xi_{\vk}&\lambda^*_{\vk} 
	\end{pmatrix}/\mathcal{N}_\vk,
\end{eqnarray}
where $\xi_{\vk}=(\xi_{s\vk}-\xi_{d\vk})/2$, $\epsilon_\vk=\sqrt{\xi_{\vk}^2+|\lambda_{\vk}|^2}$ and $\mathcal{N}_\vk=\sqrt{2\epsilon_\vk(\epsilon_\vk-\xi_{\vk})}$ is a normalization factor. The pairing matrix can then be transformed into the band basis in the following manner,
\begin{eqnarray}\label{eqn12}
	\tilde{\Delta}_\vk=\begin{pmatrix}
		\Delta_{1\vk} & \Delta_{12\vk} \\
		\Delta_{21\vk}&\Delta_{2\vk} 
	\end{pmatrix} 
	=\hat{U}_\vk^{-1}\begin{pmatrix}
		\Delta_{s} &  \\
		&\Delta_{d} 
	\end{pmatrix}\hat{U}^*_{\bar{\vk}}
	=\begin{pmatrix}
		\frac{(\epsilon_\vk+\xi_{\vk})\Delta_s+(\epsilon_\vk-\xi_{\vk})\Delta_d}{2\epsilon_\vk} & \frac{\lambda_{\vk}(\Delta_s-\Delta_d)}{2\epsilon_\vk}  \\
		\frac{\lambda_{\vk}(\Delta_s-\Delta_d)}{2\epsilon_\vk} &\frac{(\epsilon_\vk-\xi_{\vk})\Delta_s+(\epsilon_\vk+\xi_{\vk})\Delta_d}{2\epsilon_\vk} 
	\end{pmatrix} \,.
\end{eqnarray}
In the above, we have used the fact that the orbital hybridization in our model with $s$ and $d_{xy}$ orbitals satisfies the relation $\lambda_\vk = \lambda^*_{\bar{\vk}}$. It is easy to see that, if $\Delta_s=\pm\Delta_d$, then $\Delta_{1\vk}=\pm\Delta_{2\vk}$. To a good approximation, the interband pairing term $\Delta_{12\vk}$ is ignored in our weak-coupling analysis.  
\subsection{\label{sec:level2} Modeling of the $\pi$-PDW state}
The $\pi$-PDW phase is characterized by a phase modulation with a periodicity of two lattice stretching in the $x$-direction. Hence we write the kinetic part of the Hamiltonian on the basis of a unitcell with two neighboring sites $(c_{As\vk},c_{Bs\vk},c_{Ad\vk},c_{Bd\vk})^T$, where $A,B$ label the two sublattices, $s$ and $d$ denote the two electron orbitals,
\begin{eqnarray}\label{eq:HkPDW}
	\hat{H}_{0,\vk}=
	\begin{pmatrix}
		-2t_s\cos k_y -\mu_s&-2t_s\cos k_x&&-4t'\sin k_x \sin k_y \\
		-2t_s\cos k_x &-2t_s\cos k_y -\mu_s&-4t'\sin k_x \sin k_y& \\
		&-4t'\sin k_x \sin k_y&-2t_d\cos k_y -\mu_d& -2t_d\cos k_x \\
		-4t'\sin k_x \sin k_y&& -2t_d\cos k_x& -2t_d\cos k_y -\mu_d
	\end{pmatrix}
	.\end{eqnarray}
Here $k_x$ ranges from $-\pi/2$ to $\pi/2$ due to the doubling of lattice constant in this direction, and the spin indices have been suppressed. Note that in the above construction, we have chosen the gauge under which the relative position between the sublattice sites are manifested. Such a gauge is important to obtaining the correct forms of the velocity operators, as electron hoppings on all of the bonds must be accounted for on equal footing. In this gauge, the (normal state) velocity operator is straightforwardly obtained by taking the derivative of (\ref{eq:HkPDW}), i.e. $\hat{V}_{0,\mu\vk} = \partial_{k_\mu} \hat{H}_{0,\vk}$. 

The BdG Hamiltonian of the PDW phase can be written in the Nambu spinor basis $(c_{As\vk\uparrow},c_{Bs\vk\uparrow},c_{Ad\vk\uparrow},\linebreak c_{Bd\vk\uparrow},c^\dagger_{As\bar{\vk}\downarrow},c^\dagger_{Bs\bar{\vk}\downarrow},c^\dagger_{Ad\bar{\vk}\downarrow},c^\dagger_{Bd\bar{\vk}\downarrow})^T$ as,
\begin{eqnarray}
	\hat{H}_{\vk} = 	\begin{pmatrix}
		\hat{H}_{0,\vk} & \hat{\Delta}_{\vk} \\
		\hat{\Delta}_{\vk}^\dagger & -\hat{H}^T_{0,\bar{\vk}}
	\end{pmatrix}\,.
\end{eqnarray}
In our study, the pairing matrix with on-site spin-singlet intra-orbital pairing on the two respective orbitals is given by,
\begin{eqnarray}\label{eqn11}
	\hat{\Delta}_{\vk}=
	\begin{pmatrix}
		\Delta_{As} & 0 & 0 & 0 \\
		0 & \Delta_{Bs} & 0 & 0 \\
		0 & 0 & \Delta_{Ad} & 0 \\
		0 & 0& 0 & \Delta_{Bd}
	\end{pmatrix}.
\end{eqnarray}
The relative phase between the two intra-orbital pairings on each site characterizes the configuration of the pairing potentials. That is, $\text{sgn}[\Delta_{A(B)s}] = - \text{sgn}[\Delta_{A(B)d}]$ for the $\hat{\Delta}_{+-}$ configuration and $\text{sgn}[\Delta_{A(B)s}] = \text{sgn}[\Delta_{A(B)d}]$ for the $\hat{\Delta}_{++}$ configuration. On the other hand, for the $\pi$-PDW state, we have $\Delta_{A s(d)}=-\Delta_{Bs(d)}$. The same Hamiltonian can be used to simulate the uniform phase if we let $\Delta_{A s(d)}=\Delta_{Bs(d)}$.

With the interactions given in Eq.~(\ref{eq:interactions}) of the maintext, the matrix elements in (\ref{eqn11}) can be self-consistently determined by solving the following gap equations, 
\begin{eqnarray}\label{eqn9}
	\Delta_{P s}&=&\sum_{\vk} \left(-U_{ss}\langle c_{P s \bar{\vk}\downarrow}c_{P s\vk\uparrow} \rangle +U_{sd}\langle c_{P d \bar{\vk}\downarrow}c_{P d \vk\uparrow} \rangle \right), \nonumber \\
	\Delta_{P d}&=&\sum_{\vk} \left(-U_{dd}\langle c_{P d \bar{\vk}\downarrow}c_{P d \vk\uparrow} \rangle +U_{sd}\langle c_{P s \bar{\vk}\downarrow}c_{P s\vk\uparrow} \rangle \right), ~~~~~~~~~~\text{P=A, B}. \end{eqnarray}
Note that the summation over $k_x$ ranges from $-\pi/2$ to $\pi/2$ as mentioned above. 

\begin{thebibliography}{99}
	\bibitem{Thouless:82} D. J. Thouless, M. Kohmoto, M. P. Nightingale, and M. den Nijs, Phys. Rev. Lett. \textbf{49}, 405 (1982).
	\bibitem{Hasan:10} M. Z. Hasan and C. L. Kane, Rev. Mod. Phys. \textbf{82}, 3045 (2010).
	\bibitem{Qi:11} X. L. Qi and S. C. Zhang, Rev. Mod. Phys. \textbf{83}, 1057 (2011).
	\bibitem{Neupert:13} T. Neupert, C. Chamon, and C. Mudry, Phys. Rev. B \textbf{87}, 245103 (2013). 
	\bibitem{LiZ:20} Z. Li, T. Tohyama, T. Iitaka, H. Su, and H. Zeng, Sci. China Phys. Mech. Astron. \textbf{64}, 107211 (2021).
	\bibitem{Topp:21} G. E. Topp, C. J. Eckhardt, D. M. Kennes, M. A. Sentef, and P. T\"orm\"a, Phys. Rev. B \textbf{104}, 064306 (2021).
	\bibitem{Ahn:21} J. Ahn, G. Y. Guo, N. Nagaosa, and A. Vishwanath, Nat. Phys. \textbf{18}, 3 (2022).
	\bibitem{Peotta:15} S. Peotta and P. T\"orm\"a, Nat. Commun. \textbf{6}, 8944 (2015).
	\bibitem{Julku:16} A. Julku, S. Peotta, T. I. Vanhala, D. H. Kim, and P. T\"orm\"a, Phys. Rev. Lett. \textbf{117}, 045303 (2016).
	\bibitem{Liang:17} L. Liang, T. I. Vanhala, S. Peotta, T. Siro, A. Harju, and P. T\"orm\"a, Phys. Rev. B \textbf{95}, 024515 (2017).
	\bibitem{Torma:21} P. T\"orm\"a, S. Peotta, and B.A. Bernevig, Nat. Rev. Phys. \textbf{4}, 528 (2021).
	\bibitem{Huhtinen:22}  K. Huhtinen, J. Herzog-Arbeitman, A. Chew, B. A. Bernevig, and P. T\"orm\"a, Phys. Rev. B \textbf{106}, 014518 (2022)
	\bibitem{Hazra:19} T. Hazra, N. Verma, and M. Randeria, Phys. Rev. X \textbf{9}, 031049 (2019).
	\bibitem{Verma:21} N. Verma, T. Hazra, and M. Randeria, Proc. Natl. Acad. Sci. \textbf{118}, e2106744118 (2021).
	\bibitem{Hu:19} X. Hu, T. Hyart, D. I. Pikulin, and E. Rossi, Phys. Rev. Lett. \textbf{123}, 237002(2019).
	\bibitem{Julku:20} A. Julku, T. J. Peltonen, L. Liang, T. T. Heikkil\"o, and T\"orm\"a, Phys. Rev. B \textbf{101}, 060505 (2020).
	\bibitem{Xie:20} F. Xie, Z. Song, B. Lian, and B. A. Bernevig, Phys. Rev. Lett. \textbf{124}, 167002 (2020).
	\bibitem{Cao:18} Y. Cao, V. Fatemi, S. Fang, K. Watanabe, T. Taniguchi, E. Kaxiras, and P. Jarillo-Herrero, Nature \textbf{556}, 43 (2018).
	\bibitem{Yankowitz:19} M. Yankowitz, S. Chen, H. Polshyn, Y. Zhang, K. Watanabe, T. Taniguchi, D. Graf, A. F. Young, and C. R. Dean, Science \textbf{363}, 1059 (2019).
	\bibitem{Chen:21} W. Chen and W. Huang, Phys. Rev. Research \textbf{3}, L042018 (2021).
	\bibitem{WangZQ:20} Z. Wang, G. Chaudhary, Q. Chen, and K. Levin, Phys. Rev. B {\bf 102}, 184504 (2020).
	\bibitem{Ahn:21prb} J. Ahn and N. Nagaosa, Phys. Rev. B \textbf{104}, L100501 (2021).
	\bibitem{Kitamura:22a} T. Kitamura, T. Yamashita, J. Ishizuka, A. Daido, and Y. Yanase, Phys. Rev. Research \textbf{4}, 023232 (2022).
	\bibitem{Fulde:64} P. Fulde and R. A. Ferrell, Phys. Rev. \textbf{135}, A550 (1964).
	\bibitem{Larkin:65} A. I. Larkin and Y. N. Ovchinnikov, Sov. Phys. JETP \textbf{20}, 762 (1965).
	\bibitem{Agterberg:2017} D. F. Agterberg, P. M. R. Brydon, and C. Timm, Phys. Rev. Lett. \textbf{118}, 127001 (2017). 
	\bibitem{Fradkin:15} E. Fradkin, S. A. Kivelson, and J. M. Tranquada, Rev. Mod. Phys. \textbf{87}, 457 (2015).
	\bibitem{Agterberg:20}D. F. Agterberg, J. C. S. Davis, S. D. Edkins, E. Fradkin, D.
	Harlingen, S. A. Kivelson, P. A. Lee, L. Radzihovsky, J. M. Tranquada, and Y. Wang, Annu. Rev. Condens. Matter Phys. \textbf{11}, 231 (2020)
	\bibitem{Yang:21} Y. F. Jiang, H. Yao, and F. Yang, Phys. Rev. Lett. \textbf{127}, 187003 (2021).
	\bibitem{Kitamura:22} T. Kitamura, A. Daido, and Y. Yanase, Phys. Rev. B \textbf{106}, 184507 (2022). 
	\bibitem{Jiang:23} G. Jiang and Y. Barlas, Phys. Rev. Lett. \textbf{131}, 016002 (2023).
\end{thebibliography}
\end{document}